\documentclass{mpe_report}

\usepackage{psfig,graphicx,epsfig}
\usepackage{color}
\usepackage{amsmath,amssymb,epic,eepic,array}

\begin{document}

\pagenumbering{arabic}
\setcounter{page}{40}

\renewcommand{\FirstPageOfPaper }{ 40}\renewcommand{\LastPageOfPaper }{ 43}
\def\laeq{\raise.2ex\hbox{$<$}\kern-.75em\lower.9ex\hbox{$\sim$}\,}
\def\gaeq{\raise.2ex\hbox{$>$}\kern-.75em\lower.9ex\hbox{$\sim$}\,}
\hyphenation{Reich con-ti-nuum}

\title{Constraints on the Parameters of the Unseen Pulsar in the PWN G0.9+0.1 from Radio, X-Ray, and VHE Gamma-Ray Observations}
\author{C.Venter\inst{1} \and O.C. de Jager\inst{1}}  
\institute{$^1$Unit for Space Physics, School of Physics, North-West University, Potchefstroom Campus, 2520, Potchefstroom, South Africa}
\maketitle

\begin{abstract}
Radio, X-ray, and H.E.S.S.\ gamma-ray observations of the Galactic Center (GC) composite supernova remnant SNR~G0.9+0.1 are used to constrain a time-dependent injection model of the downstream electron spectrum responsible for the total multiwavelength spectrum. The effect of  spindown power evolution as well as nebular field evolution is employed to reproduce the present-day multiwavelength spectrum. Assuming a nebular magnetic field decay model of typical H.E.S.S.-type pulsar wind nebulae (PWN), ending with a present-day field strength of $6\mu$G, we obtain an initial spindown power of $\sim 10^{38}$ ergs/s if we assume a birth period and age of 43~ms and $6,500$ yr respectively to reproduce the properties of the SNR shell. This gives a present-day spindown power of $\sim 10^{37}$ ergs/s, which agrees well with the present-day spindown power derived from X-ray observations.
\end{abstract}

\section{Introduction}
The H.E.S.S.\ Collaboration recently reported the detection of the composite supernova remnant SNR~G0.9+0.1 in very high energy (VHE) gamma-rays~[\cite{HESS05}] with a significance of $\approx 13\sigma$, after 50~hours of observations of the Galactic Center (GC) region during March-September~2004 with the full telescope system (see also~[\cite{A06}]). The unresolved
gamma-ray radiation, which appears to be associated with the plerionic remnant core, corresponds to a photon flux above 200~GeV of $(5.7\pm0.7_{stat}\pm1.2_{sys})\times10^{-12}$/cm$^2$/s and luminosity of $\sim2\times10^{34}$ ergs/s, making this one of the faintest VHE gamma-ray sources. For a radially symmetric Gaussian emission region, a 95\% confidence limit of $1.3^\prime\sim3.2$~pc on the source radius is obtained, and similarly for emission from a uniform thin shell, with extension $<2.2^\prime\sim5.4$ pc. This is the first time that SNR~G0.9+0.1 has been detected at gamma-ray energies~[\cite{deJ06}], subsequent to the earlier unconstraining upper limit given by \textit{HEGRA}~[\cite{A02}], and H.E.S.S.\ observations in 2003 (resulting in a $4\sigma$ detection)~[\cite{A04}]. \textit{MAGIC} also detected a small excess from G0.9+0.1~[\cite{MAGIC06}], but it is however not statistically significant yet due to limited observation time. 

Helfand and Becker~[\cite{HB87}] observed G0.9+0.1 in 1984 for 45~minute integrations at 20~cm and 6~cm with the \textit{VLA} in their search for further examples of composite objects, despite its omission from Green's first catalogue~[\cite{G84}]. This led to the discovery of the composite nature of this bright, extended source near the GC in the radio band, with the flat radio core, corresponding to the pulsar wind nebula (PWN), and steeper shell components clearly distinguishable. (Shortly afterwards, Reich, Sofue \& Fuerst~[\cite{RSF87}] confirmed G0.9+0.1's non-thermal nature and its identification as an SNR by detailed comparison of radio continuum and infrared emission from the GC). Using various previous observations, Helfand and Becker assigned spectral indices of $\alpha_c\approx0.1$ (typical of Crab-like remnants) and $\alpha_s\approx0.6$ to the core and shell components, and also constructed a broad band spectrum (radio -- X-ray) of this source, as well as an upper limit of 3~Jy on the $25\mu m$ infrared flux.
(Recent \textit{VLA} measurements at 90~cm~[\cite{LKLH00}] revealed similar indices of 0.12 and 0.77 for the core and shell components, confirming this SNR's composite classification~[\cite{WP78}]; see~[\cite{deJV05}] for a recent review).

Helfand and Becker~[\cite{HB87}] furthermore indicated a core-to-shell diameter ratio of $\sim0.25$. Assuming a distance of 10~kpc, they obtained radio luminosities of $L_{R,c}=7.0\times 10^{34}$ ergs/s and $L_{R,s}=2.8\times10^{34}$ ergs/s for the core and shell components, and an X-ray luminosity ($0.5-4.5$~keV) of $\sim(1-5)\times10^{34}$ ergs/s for this remnant, setting the X-ray-to-radio flux ratio for the core to \laeq 0.6. Other cited luminosity ratios are $L_{R,c}/L_{R,s}\sim2.5$ and $L_{X,c}/L_{X,s}\sim0.5$. These authors also argued for an initial spin period of a few ms for the putative pulsar, but further \textit{VLA} observations at 6~cm and 20~cm did not reveal any point source above 6$\sigma$ (0.45~mJy)~[\cite{FM93}]. Observations during a continuum GC survey at 843~MHz by \textit{MOST}~[\cite{G94}] confirmed a core and shell component, with respective flux densities in close agreement with the spectra given by [\cite{HB87}]. Owing to G0.9+0.1's structure being similar to that of SNR~0540-693 (in the Large Magellanic Cloud) which contains a pulsar, G0.9+0.1 was identified as a likely candidate for being a pulsar host. High dispersion measures and interstellar scattering may however inhibit detection of pulsed emission from G0.9+0.1~[\cite{G94}].

While performing a survey of the GC, \textit{BeppoSAX} serendipitously detected SNR~G0.9+0.1 [\cite{MSI98}]. This detection represented the first firm evidence of X-ray emission from G0.9+0.1, in addition to an earlier marginal detection by the \textit{Einstein Observatory} [\cite{HB87}] and a nondetection by \textit{ROSAT} [\cite{MSI98,SBM01}]. Follow-up observations of the GC and on-axis observations of G0.9+0.1 [\cite{S98,S99,S99b,SMI00}] confirmed the non-thermal nature of X-rays from its radio core, and thus the composite nature of this remnant, and also found marginal indications of a diffuse character. Only an upper limit for X-ray emission from the SNR shell could be given, probably due to line-of-sight absorption [\cite{SWD03}]. Although no flux variations or pulsations were found, Sidoli \textit{et al.} [\cite{SMI00}] argued for the existence of a young central neutron star based on the broad band (radio \& X-ray) properties of the SNR.

The \textit{ASCA} GC survey also detected significant X-ray emission from G0.9+0.1 in the $3-10$~keV band, but no significant emission in the softer energy band [\cite{SY99}]. The emitting region was compact, in accordance with the radio core, but unresolved, and no line-like feature was observed. Similar updated results for the $0.7-10$~keV band, as well as an upper limit of $1.5^\prime$ for the apparent X-ray size, are given in [\cite{SK02}].

A higher resolution 35~ksec observation of this source in October 2000 with \textit{Chandra X-Ray Observatory} [\cite{GPG01}] revealed a clear axial symmetry for the PWN, which matched the morphology of the radio nebula, as well as a faint X-ray point source along the symmetry axis (CXOU~J174722.8-280915) which is considered to be the best candidate for emission from a central pulsar (although no pulsations were detected). A bright elliplical $5^{\prime\prime}\times8^{\prime\prime}$ clump, which may be an intermediate-latitude feature in the pulsar wind, was also observed. An equatorial torus and axial jet morphology provides a natural explanation for the data and lends support to the idea that this type of morphology might be ubiquitous within the pulsar population.

SNR~G0.9+0.1 was also easily identified in the hard energy band when a wide-angle \textit{XMM-Newton} survey of the CG Region was conducted during 2000-2001 (11 pointings of $10-25$~ksec exposures) [\cite{SWD03,S04}]. On-axis observations during September 2000 ($\sim29$ ksec) [\cite{PDW03}] revealed a large scale X-ray morphology in good correspondence with the 20~cm \textit{VLA} radio contours [\cite{HB87}], with the X-ray core matching the eastern radio peak, and the western side of the X-ray arc-like feature corresponding to the western radio peak. Diffuse emission from the radio shell was also detected in X-rays for the first time (of either thermal or non-thermal origin), in good agreement with the earlier upper limit of Sidoli \textit{et al.} [\cite{SMI00}]. Both an absorbed power-law and a thermal bremsstrahlung model provide good fits for the data in the region of the PWN. Evidence for spectral steepening with increasing radius supports a scenario where high-energy electrons undergo synchrotron losses as they diffuse through the nebula.  The spectral steepening is stronger than in other PWN (e.g.\ 3C58 \& G21.5-0.9) and the core is harder. Small-scale structure resembled that detected by \textit{Chandra}, having less detail but double the extent of the emission. Spectra of the various structures were derived for the first time, indicating that the eastern part of the arc-like feature exhibited harder X-ray emission than the western part. This may be explained in terms of relativistic beaming or Doppler boosting effects. The jet-like feature also had distinct spectral indices for its northern and southern part. Furthermore, the central pulsar candidate CXOU~J174722.8-280915 was visible above 6~keV, and magnetospheric emission seems to be preferred to a black-body model, implying a luminosity of $\sim10^{33}d^2_{10}$ ergs/s for this source ($d_{10}=d/10$ kpc). The deepest X-ray observation ever performed on this SNR [\cite{SBMB04}] confirms the spectral softening at larger radial distances from the PWN peak as well as X-ray emission from a region spatially coincident with the radio shell.

During 2003-2004, \textit{INTEGRAL} observed the GC in the $20-100$~keV energy range for an effective exposure time of $4.7\times10^6$ sec [\cite{BG06}]. Emission from a region with centroid label IGR~J17475-2822, a source associated with Sgr~B2, was clearly observed [\cite{R04}]. The extension of this source toward the north was tentatively associated with G0.9+0.1 and interpreted as a detection of VHE synchrotron radiation.

In this paper we use radio, X-ray, and gamma-ray data to constrain both a time-dependent injection model of the downstream electron spectrum (responsible for the total multiwavelength spectrum), and the evolutionary history of the spindown power.
  
\section{The Model}
  
We assume that the central pulsar's spindown power has the following functional form:
\begin{equation}
   L = I\omega\dot{\omega} = -K\omega^{n+1}\!\!\!\!\!\!\!\!\!,\label{eq:Erot}
\end{equation}
with $I$ the moment of inertia, $\omega$ the angular frequency, $\dot{\omega}$ its time-derivative, and $n = \omega\ddot{\omega}/\dot{\omega}^2$ the braking index. We next assume that $n$ is a constant (see, however, [\cite{CL06}]) and that the neutron star's crustal magnetic field does not decay. This leads to the condition
\begin{equation}
  \dot{P}P^{n-2} = \dot{P}_0P_0^{n-2}\!\!\!\!\!\!\!,\label{eq:cond}
\end{equation}
with $P = 2\pi/\omega$ the pulsar period, $\dot{P}$ its time-derivative, and the subscript `0' indicating quantities at pulsar birth. Upon integration of $\dot{\omega}$ from eq.~(\ref{eq:Erot}), and using eq.~(\ref{eq:cond}), one finds the following general expression for the evolution of $L(t)$ (see also [\cite{RC84}]):
\begin{equation}
  L(t) = L_0\left[1+\frac{(n-1)P_0^2L_0t}{4\pi^2I}\right]^{-\frac{n+1}{n-1}}\!\!\!\!\!\!\!\!\!\!\!\!\!\!.\label{eq:Lt}
\end{equation}
In what follows, we assume $P_0$ = 0.043 s, as inferred by Van der Swaluw and Wu [\cite{VW01}] using the ratio of the PWN and shell radii. We model the effect of a time-changing nebular field by 
\begin{equation}
  B(t) = \frac{B_0}{1+(t/\tau_0)^\alpha}.
\end{equation}
For the electron injection spectrum $Q(E,t)$ (number of electrons per second per energy at the shock radius $r_S$), we assume a broken power law with indices $\alpha_1$ and $\alpha_2$, and break energy $E_{\rm b}$:
$$\!\!\!\!\!\!\!\!\!\!\!\!\!\!\!\!\!\!\!\!\!\!\!\!\!\!\!\!\!\!\!\!\!\!\!\!\!\!\!\!\!\!\!\!\!\!\!Q(E,t)=\left\{\begin{array}{ll}
Q_0(t)\left(\frac{E}{E_{\rm b}}\right)^{\alpha_1} & \quad E < E_{\rm b}\\
Q_0(t)\left(\frac{E}{E_{\rm b}}\right)^{\alpha_2} & \quad E \geq E_{\rm b}
\end{array}\right..$$
We normalise $Q$ by requiring that
\begin{equation}
\int Q(E,t)EdE = \epsilon L(t),
\end{equation}
with $\epsilon \sim 0.1$ a conversion efficiency of spindown power into particle luminosity. This leads to
\begin{equation}
Q_0(t) = \frac{\epsilon L(t)}{E^2_{\rm b}}\left[\frac{\left(\alpha_1+2\right)\left(\alpha_2+2\right)}{\left(\alpha_2-\alpha_1\right)}\right]\!\!,
\end{equation}
for $\alpha_2<-2$.

\begin{table}
  \begin{tabular}{|l|c|c|}
    \hline
    \textbf{Model Parameter} & \textbf{Symbol} & \textbf{Value/Range}\\
    \hline
    Braking index & $n$ & 3\\
    B-field parameter & $\alpha$ & 0.5\\
    Present-day B-field & $B(T)$ & 6 $\mu G$\\
    Conversion efficiency & $\epsilon$ & 0.1\\
    Age & $ T $ & 6,500 yr\\
    Characteristic time scale & $\tau_0$ & 500 yr\\
    Distance & $d$ & 8.5 kpc\\
    Magnetization parameter & $\sigma$ & 0.2\\
    Moment of inertia & $I$ & $10^{45}$ g.cm$^2$ \\
    Q break energy & $ E_{\rm b} $ & $10^{-8}-10$ ergs\\
    Q index 1 & $ \alpha_1$ & -1.0\\
    Q index 2 & $ \alpha_2$ & -2.2\\
    Initial spindown power & $L_0$ & $10^{36}$ -- $10^{40}$ ergs/s\\
    Birth period & $P_0$ & 0.043 s\\
    \hline
  \end{tabular}
\caption{Values of parameters used for the model (see text for details).}
\label{tab3}
\end{table}

We next impose two boundary conditions. The first is that $r_L \laeq 0.5r_S$, with  $r_L$ the electron's Larmor radius [\cite{deJ07}]. This ensures particle confinement within the PWN, and leads to a condition on the particle energy:
\begin{equation}
  E(t) < \frac{e}{2}\sqrt{\frac{\sigma L(t)}{(1+\sigma)c}},\label{eq:bcond1}
\end{equation}
with $\sigma\equiv L_{\rm EB}/L_{\rm E}$ the ratio of the electromagnetic energy flux to the particle energy flux. The second energy condition is required to ensure that the particles radiating synchrotron emission will survive until time $t=T$ (with $T$ the PWN's age):
\begin{equation}
  E(t) < \frac{422}{B(t)B(T)(T-t)}\quad{\rm ergs}.\label{eq:bcond2}
\end{equation}
We lastly calculate the leptonic spectrum $dN/dE$ subject to eq.~(\ref{eq:bcond1}) and eq.~(\ref{eq:bcond2}) by integrating $Q$ over time from $t = 0$ to $t=T=6$,500~yr (see [\cite{MSI98}], where an age of $\sim$6,800~yr is derived assuming that G09.+0.1 is expanding adiabatically). Table~\ref{tab3} summarizes the values adopted for certain model parameters.\\

\section{Results and Conclusions}
Using this time-dependent model for the nebular averaged leptonic spectrum, we add the contributions from all epochs which survive to the present-day to give the net present-day photon spectra as calculated for synchrotron and inverse Compton emission. In the latter case the target photon fields are the CMBR, as well as galactic target photon fields from the 25K dust and starlight photon fields, assuming associated energy densities of $\sim 1$ eV/cm$^3$ for each component.

\begin{figure}
\centerline{\psfig{file=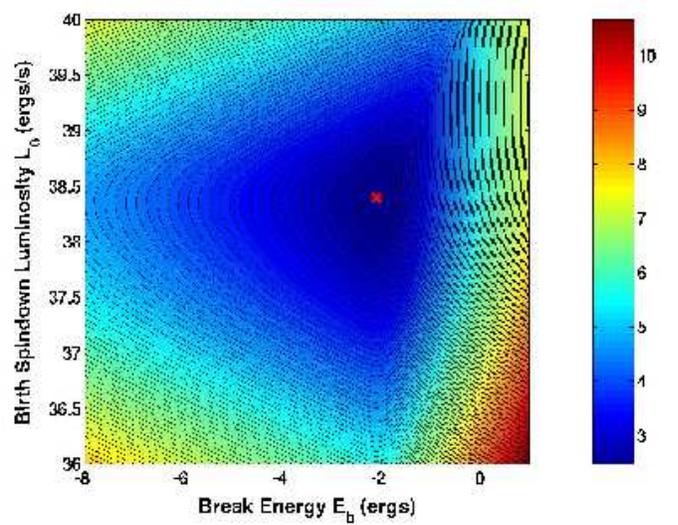,width=8.8cm,clip=}}
\caption{Plot of $\log_{10}\chi_{\rm tot}$ (with $\chi_{\rm tot}$ the rms-total of the individual test statistics for fits to the radio, X-ray and gamma-ray data) vs.\ $\log_{10}$ of initial spindown luminosity $L_0$ and $\log_{10}$ of injection spectrum break energy $E_{\rm b}$, in order to find the `best fit' (indicated by an `X') for $L_0$ and $E_{\rm b}$ (at a fixed age $T$ = 6,500~yr and birth period $P_0$ = 0.043~s).
\label{fig1}}
\end{figure}

\begin{figure}
\centerline{\psfig{file=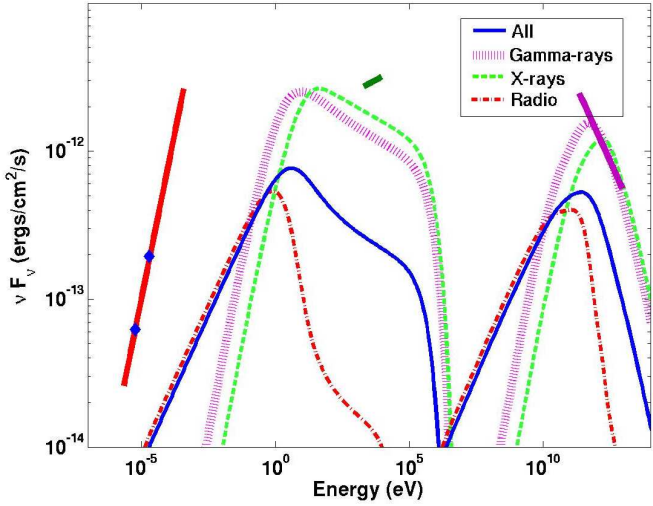,width=8.8cm,clip=}}
\caption{Plots of best fits of the spectral energy distribution (SED) of G0.9+0.1, with the different fits corresponding to Table~\ref{tab4}, i.e.\ for individual and collective fits of the multiwavelength data (at a fixed age $T$ = 6,500~yr and birth period $P_0$ = 0.043~s). Also shown are radio data points (the diamonds) and radio spectrum taken from [\cite{HB87}], X-ray spectrum from [\cite{SBMB04}], and gamma-ray spectrum from [\cite{HESS05}].
\label{fig2}}
\end{figure}

From Figure~\ref{fig1} we find an optimum fit of initial spindown luminosity $L_0\sim 2.5\times10^{38}$ ergs/s and a spectral break energy of $E_{\rm b}\sim 8\times10^{-3}$ ergs. Note that we obtain different optimum values when we fit the radio, X-ray and VHE gamma-ray data individually. The resulting fits (and multiwavelength observations) are shown in Figure~\ref{fig2} and Table~\ref{tab4}, where we also show the corresponding total energy output $E_{\rm tot}=\int L(t)dt$, integral X-ray flux $F_{\rm X}$ ($2-10$ keV) in ergs/cm$^2$/s, and integral $\gamma$-ray flux above 0.2~TeV, $F_{\gamma}(>0.2)$, in /cm$^2$/s. A distance of $d$ = 8.5 kpc is assumed [\cite{HESS05}]. These values correspond quite well to inferred values given in [\cite{HESS05,MSI98,SMI00,PDW03,SBMB04}]). 

Eq.~(\ref{eq:Lt}) now implies a value of present-day spindown power of $L(T)\sim7.4\times 10^{36}$ ergs/s, which agrees within $\sim50$\% with the inferred value of $L(T)\sim 1.5\times 10^{37}$ ergs/s from {\it BeppoSAX} observations [\cite{SMI00}]. One may then also infer a pulsar  polar cap (PC) magnetic field strength of $B_{\rm PC}\sim3.2\times10^{19}\sqrt{P_0\dot{P}_0}\sim5\times10^{12}$ G, which is a typical value for the canonical pulsar population.

We are, however, not yet satisfied with the quality of the total fit, since it is obvious that the radio data are not fitted very well. This is also the case for an alternative nebular field which is initially larger, and falls exponentially to the present-day assumed value of $B(T) = 6\mu$G. Future investigations will relax the initial spin period $P_0$ and age $T$ (keeping them free), and include extra components in the injection spectrum, to see if we can obtain better quality fits for the spectral energy distribution (SED) of G0.9+0.1.

\begin{table}
 \begin{tabular}{|l|c|c|c|c|c|c|}
    \hline
    \textbf{Data} & $E_{\rm b}$ & $L_0$ & $F_{\gamma}(>0.2)$ & $F_{\rm X}$ & $E_{\rm tot}$ & $\chi$\\
    \hline
    $\gamma$-rays & 0.5    & 38.0 & -11.4 & --    & 48.8 & -1.0\\
    X-rays        & 1.0    & 37.7 & --    & -11.6 & 48.7 & -0.35\\
    Radio         & -2.1   & 39.8 & --    & --    & 49.0 &  2.2\\
    All           & -2.1   & 38.4 & -11.9 & -12.4 & 48.9 &  2.4\\
    Obs.          & --     & --   & -11.2 & -11.2 & --   & --\\
    \hline
  \end{tabular}
\caption{Table of `best fit' parameters for individual and collective multiwavelength fits, as well as observed values [\cite{HESS05,PDW03}]. Column units are $\log_{10}$ of ergs, ergs/s, /cm$^2\!/$s, ergs/cm$^2\!/$s, ergs, and $\log_{10}$ of a dimensionless test statistic. (See text for details).}
\label{tab4}
\end{table}
  
\begin{acknowledgements}
This publication is based upon work supported by the South African National Research Foundation under
Grant number 2053475.
\end{acknowledgements}
   


      \clearpage


\begin{thebibliography}{} 
\bibitem[Aharonian et al. 2002]{A02}~Aharonian, F.\ \textit{et al.}, 2002, A\&A, 395, 803
\bibitem[Aharonian et al. 2004]{A04}~Aharonian, F.\ \textit{et al.}, 2004, A\&A, 425, L13
\bibitem[Aharonian et al. 2005]{HESS05}~Aharonian, F.\ \textit{et al.}, 2005, A\&A, 432, L25
\bibitem[Aharonian et al. 2006]{A06}~Aharonian, F.\ \textit{et al.}, 2006, ApJ, 636, 777
\bibitem[Albert et al. 2006]{MAGIC06}~Albert, J.\ \textit{et al.}, 2006, ApJL, 638, L101
\bibitem[B{\'e}langer et al. 2006]{BG06}~B{\'e}langer, G.\ \textit{et al.}, 2006, ApJ, 636, 275
\bibitem[Chen \& Li 2006]{CL06}~Chen, W.C., \& Li, X.D., 2006, A\&A, 450, L1
\bibitem{deJ06}  [de Jager 2005]~de Jager, O.C., 2005, in Astrophysical Sources of High Energy Particles and Radiation, ed.\ T. Bulik, B. Rudak, \& G. Madejski, AIP Conf.\ Proc., 801, 298.
\bibitem{deJ07}  [de Jager 2007]~de Jager, O.C., 2007, to appear in Neutron Stars and Pulsars, 40 Years After the Discovery, ed.\ W. Becker
\bibitem{deJV05} [de Jager \& Venter 2005]~de Jager, O.C., \& Venter, C. 2005, in Towards a Network of Atmospheric Cherenkov Detectors VII, ed.\ B. Degrange, G. Fontaine (astro-ph/0511098)
\bibitem{FM93}   [Frail \& Moffett 1993]~Frail, D.A., \& Moffett, D.A., 1993, ApJ, 408, 637
\bibitem{GPG01}  [Gaensler et al. 2001]~Gaensler, B.M., Pivovaroff, M.J., \& Garmire, G.P. 2001, ApJ, 556, L107
\bibitem{G94}    [Gray 1994]~Gray, A.D., 1994, MNRAS, 270, 835
\bibitem{G84}    [Green 1984]~Green, D.A., 1984, MNRAS, 209, 449
\bibitem{HB87}   [Helfand \& Becker 1987]~Helfand, D.J., \& Becker, R.H., 1987, ApJ, 314, 203
\bibitem{LKLH00} [LaRosa et al. 2000]~LaRosa, T.N., Kassim, N.E., Lazio, T.J.W. \& Hyman, S.D., 2000, ApJ, 119, 207
\bibitem{MSI98}  [Mereghetti et al. 1998]~Mereghetti, S., Sidoli, L., \& Israel, G.L., 1998, ApJ, 331, L77
\bibitem{PDW03}  [Porquet et al. 2003]~Porquet, D., Decourchelle, A., \& Warwick, R.S., 2003, A\&A, 401, 197
\bibitem{RSF87}  [Reich et al. 1987]~Reich, W., Sofue, Y., \& Fuerst, E., 1987, Publ.\ Astron.\ Soc.\ of Japan, 39, 573
\bibitem{R04}    [Revnivtsev et al. 2004]~Revnivtsev, M.G.\ \textit{et al.}, 2004, A\&A, 425, L49
\bibitem{RC84}   [Reynolds \& Chevalier 1984]~Reynolds, S.P, \& Chevalier, R.A., 1984, ApJ, 278, 630 
\bibitem{SK02}   [Sakano et al. 2002]~Sakano, M., Koyama, K., Murakami, H., Maeda, Y., \& Yamauchi, S., 2002,
ApJS, 138, 19
\bibitem{SWD03}  [Sakano et al. 2003]~Sakano, M., Warwick, R.S., \& Decourchelle, A., 2003, in Workshop on Galaxies and Clusters of Galaxies, 9
\bibitem{S04}    [Sakano et al. 2004]~Sakano, M., Warwick, R.S., Hands, A. \& Decourchelle, A., 2004, Mem.S.A.It., 75, 498
\bibitem{SY99}   [Sakano et al. 1999]~Sakano, M., Yokogawa, J., \& Murakami, H., 1999, in Japanese-German Workshop on High Energy Astrophysics, ed.\ W.\ Becker, \& M. Itoh, 113
\bibitem{S98}    [Sidoli et al. 1998]~Sidoli, L. \textit{et al.}, 1998, in The Active X-ray Sky: Results from \textit{BeppoSAX} and \textit{RXTE} (Nucl.\ Phys.\ B Proc.\ Suppl.\ 69), ed. L. Scarsi, H. Bradt, P. Giommi, \& F. Fiore (Amsterdam: Elsevier), 88
\bibitem{S99}    [Sidoli et al. 1999a]~Sidoli, L. \textit{et al.}, 1999,  Astroph.\ Lett.\ Comm., 38, 309 (astro-ph/9812349)
\bibitem{SBM01}  [Sidoli et al. 2001]~Sidoli, L., Belloni, T., \& Mereghetti, S., 2001, A\&A, 368, 835
\bibitem{SBMB04} [Sidoli et al. 2004]~Sidoli, L., Bocchino, F., Mereghetti, S., \& Bandiera, R., 2004, Mem.S.A.It., 75, 507
\bibitem{S99b}   [Sidoli et al. 1999b]~Sidoli, L., Mereghetti, S., Israel, G.L., Chiappetti, L., Treves, A., \& Orlandini, M., 1999, ApJ, 525, 215
\bibitem{SMI00}  [Sidoli et al. 2000]~Sidoli, L., Mereghetti, S., Israel, G.L., \& Bocchino, F., 2000, A\&A, 361, 719
\bibitem{VW01}   [Van der Swaluw \& Wu 2001]~Van der Swaluw, E., \& Wu, Y., 2001, ApJ, 555, L49
\bibitem{WP78}   [Weiler \& Panagia 1978]~Weiler, K.W., \& Panagia, N. 1978, A\&A, 70, 419
\end{thebibliography}
\end{document}